# Detection of Water and/or Hydroxyl on Asteroid (16) Psyche

Driss Takir[1], Vishnu Reddy[2], Juan A. Sanchez[3], Michael K. Shepard[4], Joshua P. Emery[5]

[1]U.S. Geological Survey, Astrogeology Science Center, Flagstaff, AZ 86001, USA (dtakir@usgs.gov)
[2]Lunar and Planetary Laboratory, University of Arizona, Tucson, AZ 85721, USA
[3]Planetary Science Institute, 1700 E Fort Lowell Road, 8 Suite 106, Tucson, AZ 85719, USA
[4]Department of Geography and Geosciences, Bloomsburg University of Pennsylvania, 400 E. Second St., Bloomsburg, PA 17815, USA
[5]Earth and Planetary Science Department, Planetary Geosciences Institute, University of Tennessee, Knoxville, TN 37996, USA

**Abstract:**

In order to search for evidence of hydration on M-type asteroid (16) Psyche, we observed this object in the 3-µm spectral region using the long-wavelength cross-dispersed (LXD: 1.9-4.2 µm) mode of the SpeX spectrograph/imager at the NASA Infrared Telescope Facility (IRTF). Our observations show that Psyche exhibits a 3-µm absorption feature, attributed to water or hydroxyl. The 3-µm absorption feature is consistent with the hydration features found on the surfaces of water-rich asteroids, attributed to OH- and/or $H_2O$-bearing phases (phyllosilicates). The detection of a 3-µm hydration absorption band on Psyche suggests that this asteroid may not be metallic core, or it could be a metallic core that has been impacted by carbonaceous material over the past 4.5 Gyr. Our results also indicate rotational spectral variations, which we suggest reflect heterogeneity in the metal/silicate ratio on the surface of Psyche.

**Introduction:**

(16) Psyche, an M-type asteroid (Tholen 1984), is thought to be one of the most massive exposed iron metal bodies in the asteroid belt (Bell et al. 1989). Bottke et al. (2006) find that iron-meteorite parent bodies, including Psyche, are most probably formed in the terrestrial planet region. Recent polarimetric and radar estimates suggest that Psyche has an irregular shape with a diameter of ~ 200 km (Lupishko 2006; Shepard et al. 2008). The high radar albedos of Psyche suggest that it is dominantly composed of metal (almost entirely Fe-Ni) (Shepard et al. 2008, 2016). Psyche's high thermal inertia (133 or 114 ± 40 J $m^{-2}$ $S^{-0.5}$ $K^{-1}$) (Matter et al. 2013) and density (6.98 ± 0.58 g $cm^{-3}$) (Kuzmanoski & Koračević 2002) provide more evidence that this asteroid has a metal-rich surface.

M-type asteroids are part of the spectroscopic X-complex taxonomy with featureless and moderately red spectra in the visible and near-infrared (NIR) region (0.3-2.5-µm) (Bus and Binzel 2002; DeMeo et al. 2009). However, Hardersen et al. (2005) and Fornasier et al. (2010) found weak 0.9 µm and 0.43 µm absorption features on several M-asteroids, including Psyche, indicating the presence of anhydrous (e.g., orthopyroxene) and hydrous (e.g., serpentine) minerals on their surfaces. An absorption feature at 1.9 µm, possibly attributed to mafic minerals, was also detected in several M-type asteroids (e.g., Hardersen et al. 2011, Ockert-Bell et al. 2010, Neeley et al. 2014). Recent NIR observations by Sanchez et al. (2016) revealed rotational spectral variations indicating a possible change in the metal/silicate ratio on the surface of Psyche.

Here we present the first comprehensive study to detect and interpret the 3-µm absorption feature on Psyche as a function of rotation phase. In a companion paper (Sanchez et al. 2016), these data are combined with rotationally resolved NIR spectra and radar observations (Shepard et al. 2016) of Psyche to provide a more comprehensive view of the asteroid.

## 2. Methodology:

*2.1. Observational techniques*

Spectra of Psyche were measured using the long-wavelength cross-dispersed (LXD: 1.9-4.2 µm) mode of the SpeX spectrograph/imager at the NASA Infrared Telescope Facility (IRTF). LXD data were obtained during two different nights in December 2015. Observational circumstances are presented in Table 1. SpeX has two detectors: a Teledyne 2048x2048 Hawaii-2RG array for the spectrograph and a science grade 1024x1024 Aladdin 3 InSb array that images the slit (Rayner et al., 2003, 2004). We were able to obtain K-band (1.95-2.5 µm) and L-band (2.85-4.1 µm) data simultaneously, using the LXD mode of SpeX and an 0.8 x 15 arcsec slit with an image scale of 0.15 arcsec/pixel. The LXD mode has a resolving power ($R = \lambda / \Delta\lambda$) of ~1000 and is covered in six orders, which are recorded simultaneously on the detector and stacked on the chip to get relatively broad wavelength coverage (Rayner et al. 2003). These orders were extracted separately and spliced together in the reduction process.

**Table 1.** Observational circumstances and 3-µm band parameters. The columns in this table are Date (UTC), the standard star used in the observations, Mid UTC, phase angle, V-magnitude, body centered longitude, and rotation phase.

| Observation Date | Std. Star | Mid. UTC | Phase Angle (°) | Mag. (V) | Long. (°) | Rot. Phase (°) |
|---|---|---|---|---|---|---|
| Dec 08, 2015 (set 1) | SAO 94309 | 11:08:44 | 1.8 | 9.4 | 292 | 68 |
| Dec 08, 2015 (set 2) | SAO 94309 | 12:15:15 | 1.8 | 9.4 | 197 | 163 |
| Dec 08, 2015 (set 3) | SAO 94309 | 13:20:44 | 1.8 | 9.4 | 103 | 257 |
| Dec 09, 2015 (set 4) | SAO 93936 | 14:14:40 | 1.7 | 9.4 | 127 | 23 |

*2.2. Data reduction*

To reduce Psyche's 3-µm data, we used the IDL (Interactive Data Language)-based spectral reduction tool Spextool (v4.0) (Cushing 2004). To remove the background sky (mostly OH line emission through most of the wavelength range and thermal emission from the sky and telescope longward of ~2.3 µm), we subtracted spectra of Psyche and a standard star at beam position A from spectrum at beam B. We extracted spectra by summing the flux at each channel within a user-defined aperture. Asteroid spectra were shifted to sub-pixel accuracy to align with the calibration standard star spectra and then were divided by appropriate calibration star spectra at the same airmass (±0.05) to remove telluric water vapor absorption features. LXD spectra were smoothed and binned to lower spectral resolution in order to improve the signal to noise ratio. We conducted wavelength calibration at $\lambda > 2.5$ µm using telluric absorption lines.

*2.3. Thermal Tail Removal*

We removed the thermal excess in Psyche's spectra using the methodology described in Takir and Emery (2012) and references therein. The thermal excess $\gamma_\lambda$ is defined as a measure of the thermal flux found at these wavelengths by:

$$\gamma_\lambda = \frac{R_\lambda + T_\lambda}{R_\lambda} - 1, \qquad (1)$$

where $R_\lambda$ is the reflected flux at a wavelength λ, $T_\lambda$ is the thermal flux at a given wavelength, and the quantity $R_\lambda + T_\lambda$ is the measured relative spectrum. To constrain Psyche's model thermal flux, we fitted the measured thermal excess with a model thermal excess. Then, we subtracted this model thermal flux from the measured relative spectrum of Psyche.

To calculate the thermal flux in the 3-μm region, we used the Near-Earth Asteroid Thermal Model (NEATM; Harris 1998), which is based on the Standard Thermal Model (STM) of Lebofsky et al. (1986). We obtained asteroid Psyche inputs, including heliocentric and geocentric distances, geometric albedo, and phase angle at time of observation from the Jet Propulsion Laboratory Horizon online ephemeris generator. We used the value of 0.15 as the default value for the slope parameter *G* (Bowell et al. 1989). This G value is also consistent with the published value of G ($0.12^{+0.04}_{-0.03}$) for Psyche in the Asteroid Absolute Magnitude and Slope (AAMS) Catalog (Oszkiewicz et al., 2011, Muinonen et al., 2010). Figure 1a shows the effect of changing the beaming parameter thermal corrections in Psyche's spectra. Figure 1b shows the chosen thermal model (η = 1.4, $p_v$ = 0.12) to thermally correct spectra of Psyche.

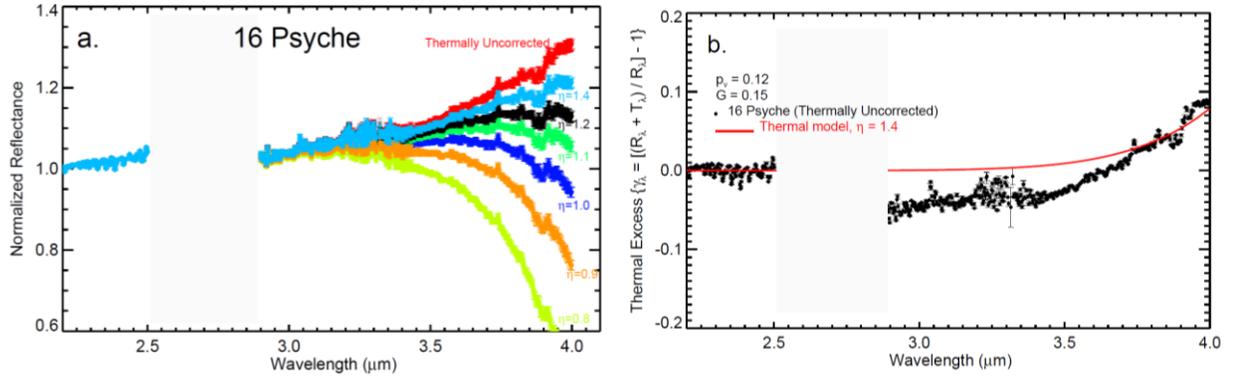

**Figure 1. (a)** The spectrum of asteroid Psyche, uncorrected (red) and corrected using thermal models with changing beaming parameter (η) values (0.8-1.4). **(b)** The best fit thermal model (in red with a beaming parameter of η = 1.4) to thermally correct spectra of Psyche.

*2.4. Continuum slope choice*

The thermal excess removal technique described in section 2.3 requires an estimate of the reflected continuum in the thermal regime. The nominal approach is to fit a regression line to the K-band portion of the spectrum and extend it through L-band. However, the slope of this reflectance continuum directly affects the band depth estimate. This dependence is not because of the thermal correction; as Figure 1 shows, the thermal correction is minimal in the 2.9-3.0 um region, regardless of thermal model parameters. However, the slope of the reflectance continuum sets the reflectance from which the band depth is measured.

In order to assess the contribution of continuum slope uncertainty to our overall band depth uncertainty, we varied the slope of regression line and shifted it vertically, then redid the thermal correction described in section 2.3. In all cases, varying the slope by 0.02 µm$^{-1}$ up or down under- or over-corrects the thermal tail a noticeable amount, so we consider that to be the reasonably possible range for the reflectance continuum. Those changes in slope change the band depth by ~1%. Therefore, a 1% uncertainty was added to the uncertainty of our calculated 3-um band depth (Table 2). An example of the thermal excess removal for different reflectance continuum slopes is shown in Figure 2 for data set 2 (Dec 8, 2015).

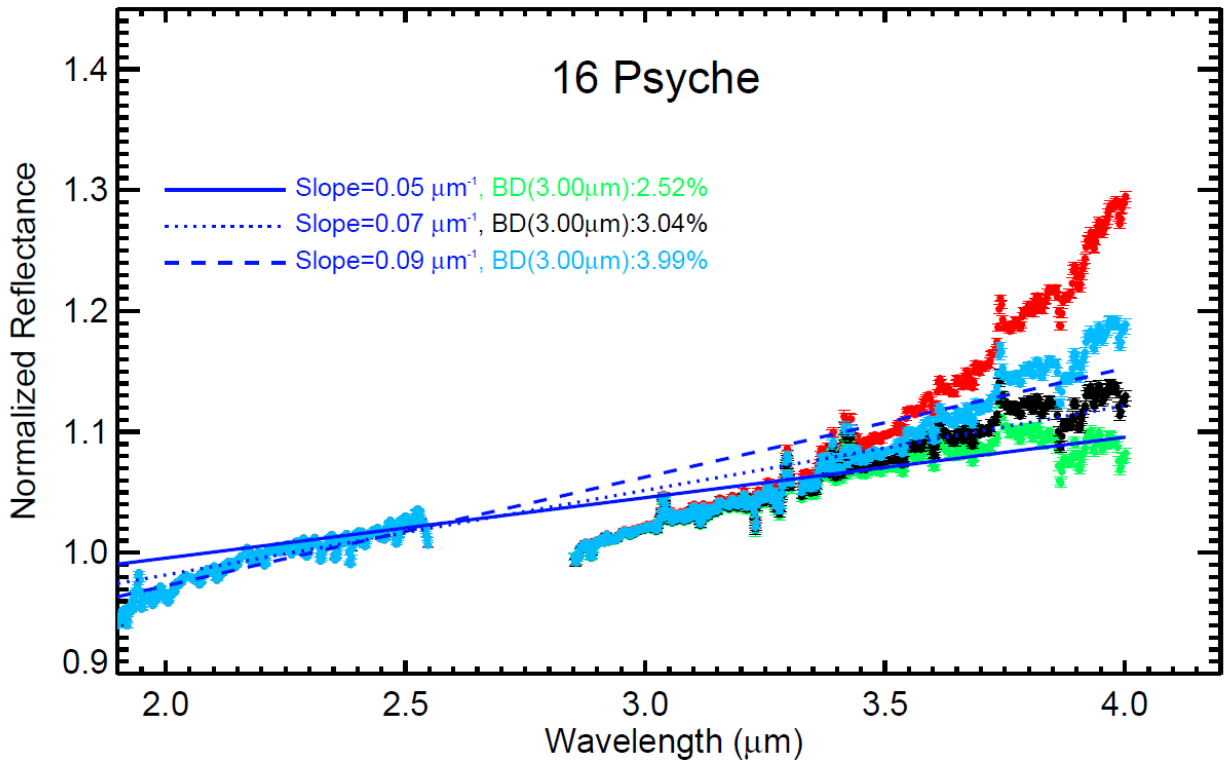

**Figure 2.** Illustration of the effect of the continuum choice on the measured band depth. The uncorrected spectrum of set 2 is shown in red. The other three spectra are the optimally corrected spectra (following the procedure described in section 2.3) for each of three choices of continuum slope, as labeled on the figure. The slope of 0.07 µm$^{-1}$ is the best fit to the data. For slopes steeper than 0.09 µm$^{-1}$, all thermal models under-correct the thermal excess, and for slopes shallower than 0.05 µm$^{-1}$, all thermal models over-correct the thermal excess.

*2.5. 3-μm Band Depth and Uncertainty*

We calculated the band depth, $D_{3.00}$, at 3 μm relative to the continuum (the linear regression line across the K-band) using the following equation:

$$D_{3.00} = \frac{R_c - R_{3.00}}{R_c},$$

(2)

where $R_{3.00}$ represents the reflectance at 3 μm, and $R_c$ represents the reflectance of the continuum at the same wavelength as $R_{3.00}$.

The uncertainty in $D_\lambda$ (the criterion is a 2σ detection) was computed using the following equations (Taylor 1982):

$$\delta D_{3.00} = D_{3.00} * \sqrt{(\frac{\delta R_1}{R_1})^2 + (\frac{\delta R_c}{R_c})^2},$$

(3)

where $\delta R_1 = \sqrt{(\delta R_c)^2 + (\delta R_{3.00})^2}$.

We also added a 1% uncertainty in quadrature to the computed uncertainties ($\sqrt{\delta D_{3.00}^2 + 0.01^2}$) to account for the uncertainty in the L-band continuum slope in all spectra.

## 3. Results:

Figure 3a-d show the LXD spectra of asteroid (16) Psyche, which were obtained during two nights in December 2015. All these spectra show absorption features around 3 μm region. Table 2 shows our 3-μm band analysis.

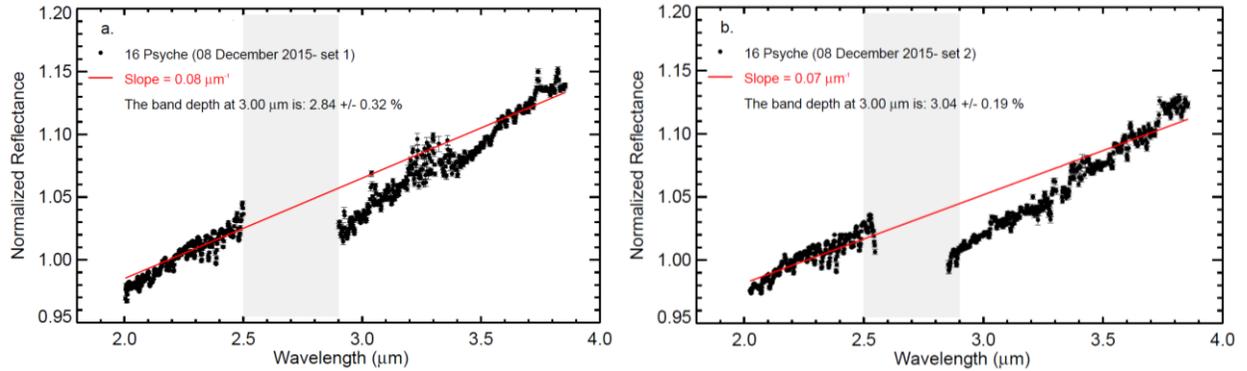
5

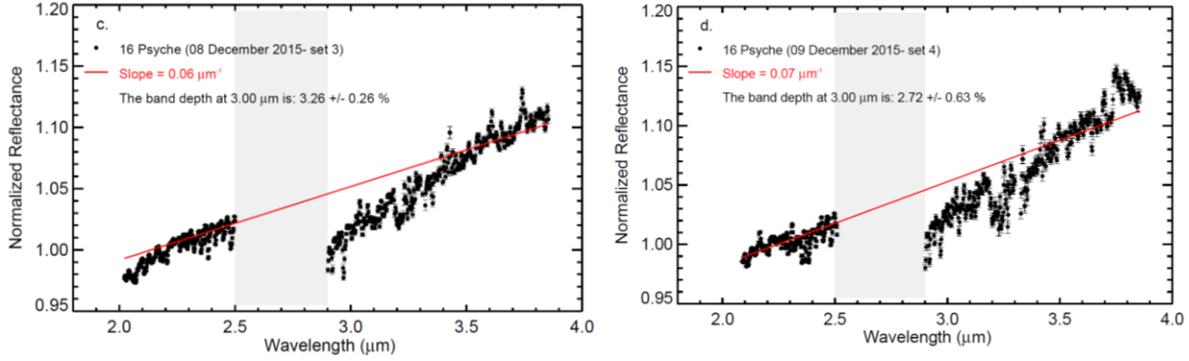

**Figure 3a-d.** LXD spectra of asteroid (16) Psyche obtained during two nights in December 2015. These spectra show 3-µm absorption features. The gray bars on each plot mark wavelengths of strong absorption by water vapor in Earth's atmosphere. The band depths were calculated using the methods described in sections 2.4 and 2.5.

**Table 2.** 3-µm band analysis showing spectra of Psyche to exhibit an absorption feature around 3 µm spectral region. A 1% uncertainty was added to our calculated 3-µm band depth uncertainty as described in *section 2.5*.

| Observation Date | Mid. UTC | 3- µm Band Depth (%) |
|---|---|---|
| Dec 08, 2015 (set 1) | 11:08:44 | 2.84±1.37 |
| Dec 08, 2015 (set 2) | 12:15:15 | 3.04±1.21 |
| Dec 08, 2015 (set 3) | 13:20:44 | 3.26±1.29 |
| Dec 09, 2015 (set 4) | 14:14:40 | 2.72±1.81 |

LXD spectra of Psyche corresponding to four different rotation phases are shown in Figure 4. These data have been combined with the average NIR spectrum of Psyche obtained by Sanchez et al. (2016) with the IRTF, so we cover the full wavelength range from ~ 0.7 to 4.2 µm. For each rotation phase, shape models of Psyche are also shown. The three-dimensional shape models of Psyche were created with the SHAPE software package (Magri et al. 2007), using data obtained by Shepard et al. (2016), which include Arecibo delay-Doppler radar observations and adaptive optics images from Keck and Magellan. A detailed description of the procedure can be found in Shepard et al. (2016) and Sanchez et al. (2016).

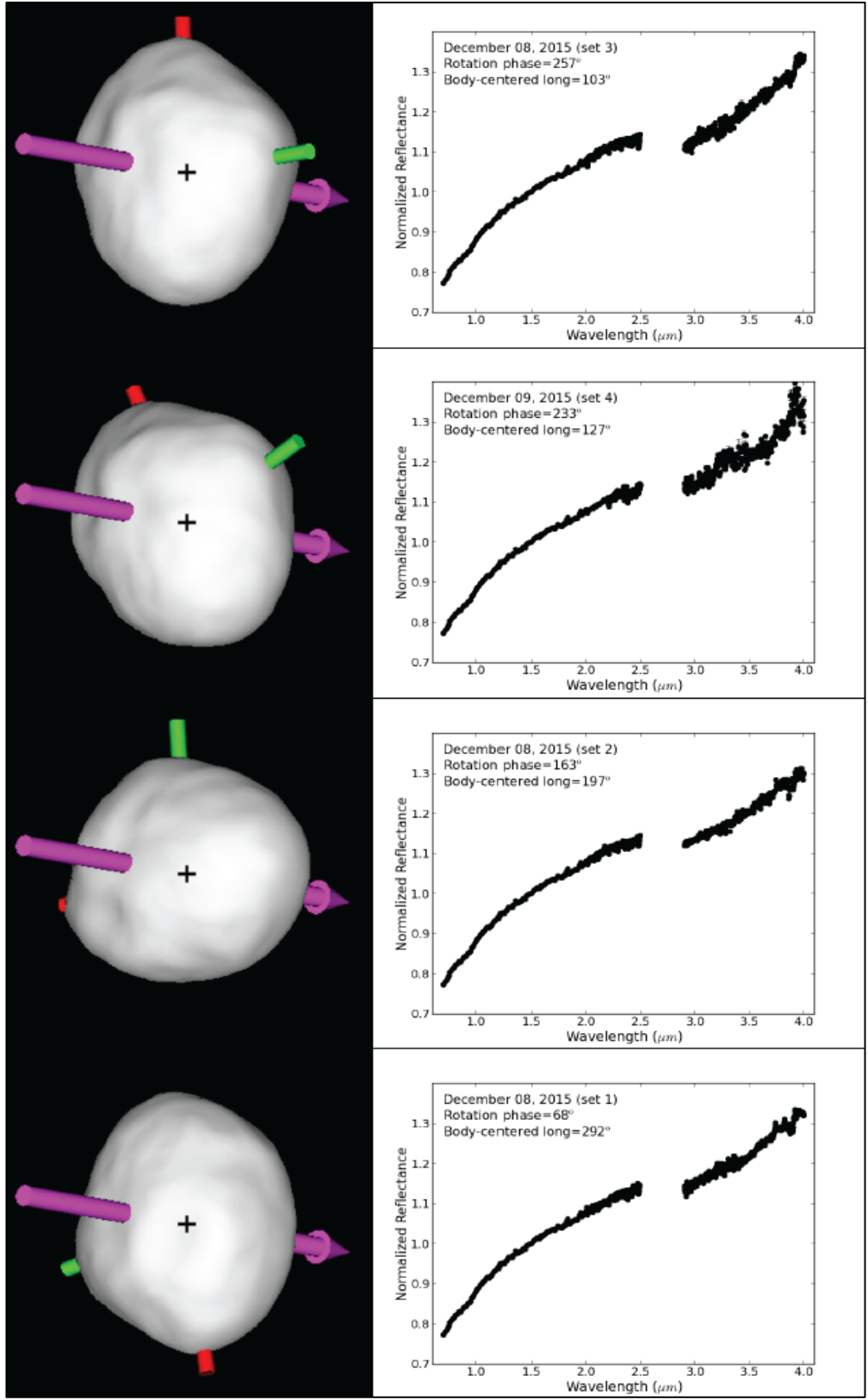

**Figure 4.** LXD spectra of (16) Psyche corresponding to four different rotation phases. These data have been combined with the average NIR spectrum of Psyche obtained by Sanchez et al. (2016). All spectra are normalized to unity at 1.5 µm. For each rotation phase, three-dimensional shape models of Psyche are also shown. The axes on the shape models correspond to longitude 0° (red) and longitude 90° (green).

## 4. Discussion

Because asteroid spectra measured with ground-based telescopes are affected by strong absorptions by water vapor in Earth's atmosphere, our analysis of Psyche LXD spectra is restricted to the 2.8-3.3-µm spectral range and the shape of the 3-µm feature. our 3-µm observational results revealed that Psyche may not be as featureless as once thought in the 3-µm spectral region. Because asteroid spectra measured with ground-based telescopes are affected by strong absorptions by water vapor in Earth's atmosphere, our analysis of Psyche LXD spectra is restricted to the 2.8-3.3-µm spectral range and the shape of the 3-µm feature.

The LXD observations have revealed that Psyche exhibits a 3-µm hydration absorption feature, similar to the hydration features observed on some water-rich asteroids (e.g., Takir and Emery 2012), attributed to OH- and/or $H_2O$-bearing phases (e.g., Takir et al. 2013). The 3-µm band shape in Psyche's spectrum is also consistent with the 3-µm band found in spectra of the Moon (Sunshine et al. 2009, Clark et al. 2009, Pieteres et al. 2009), attributed to adsorbed $H_2O$/OH.

Figure 5 shows there are some similarities between spectra of Psyche and the CI carbonaceous chondrite Ivuna (measured under asteroid-like conditions) in that they both have the same sharp 3-µm band shape (Takir et al. 2013). This sharp shape was also found in many outer Main Belt and hydrated asteroids (Takir and Emery 2012, Takir et al. 2015a), attributed to phyllosilicates (Figure 5). Figure 5 also shows that both spectra of Ivuna and the T-type asteroid (308) Polyxo have similar 3-µm band shape to the spectrum of Psyche with a deeper band depth (more abundance of hydrated minerals).

Rivkin et al. (2000) did not report a 3-µm feature in Psyche. However, given that the uncertainties in the Rivkin et al. spectrophotometry are larger than the ~3% feature observed in this work, our results are still consistent with those of Rivkin et al.

In just one set, set 1 (Figure 3a), the 3-µm feature in Psyche is superimposed on a broader absorption from ~3.25 to 3.73 µm, similarly to the feature observed on Ceres (Lebofsky 1978, Jones et al. 1990) and on the Ceres-like group (Takir and Emery 2012, Takir et al. 2015b). Milliken and Rivkin (2009) attributed these absorption features on Ceres to OH or $H_2O$-bearing phases. We only see this feature in set 1 (Dec 08, 2015), which corresponds to body-centered longitude of 292°, where a large crater (longitude 300°) of 85 ± 20km in diameter is located (Shepard et al. 2016). This impact crater on Psyche may have been a product of collisions with carbonaceous chondrite projectiles, forming a scattering of carbonaceous ejecta on its surface. All spectra show features around 3.9 µm, which is difficult to interpret because of telluric absorptions (e.g.$NO_2$) in this spectral region.

Evidence for the 3-µm band was detected in spectra of many M-type asteroids, attributed to OH or $H_2O$-bearing phases (e.g., Jones et al. 1990, Rivkin 2000, Landsman et al. 2015). Gaffey et al. (2002) and references therein proposed a number of plausible alternative interpretations for the presence of the 3-µm band on M-asteroids. These interpretations include the presence of anhydrous silicates containing structural OH, the presence of fluid inclusions, the presence of xenolithic hydrous meteorite components on asteroid surfaces from impacts (as those seen on Vesta by Reddy et al. 2012), solar wind-implanted H, or the presence of troilite.

Rivkin et al. (2000) found that M-asteroids with diameters larger than 65 km are likely to be hydrated and M-asteroids with diameters smaller than 65 km are likely to be anhydrous. The authors suggested that hydrated M-asteroids are likely to have a mineralogy similar to hydrated enstatite chondrites or high-albedo carbonaceous chondrites. In our observations we found Psyche, a large M-asteroid with a diameter of ~200 km, to be hydrated, in agreement with the work of Rivkin et al. (2000). Our observations of Psyche are also consistent with the findings of Davis et al. (1998) who suggested that large M-asteroids, like Psyche, should not be expected to be exposed metallic cores. The authors suggested that Psyche has been shattered by impacts but is not a catastrophically disrupted metallic core (e.g., Asphaug et al. 2006).

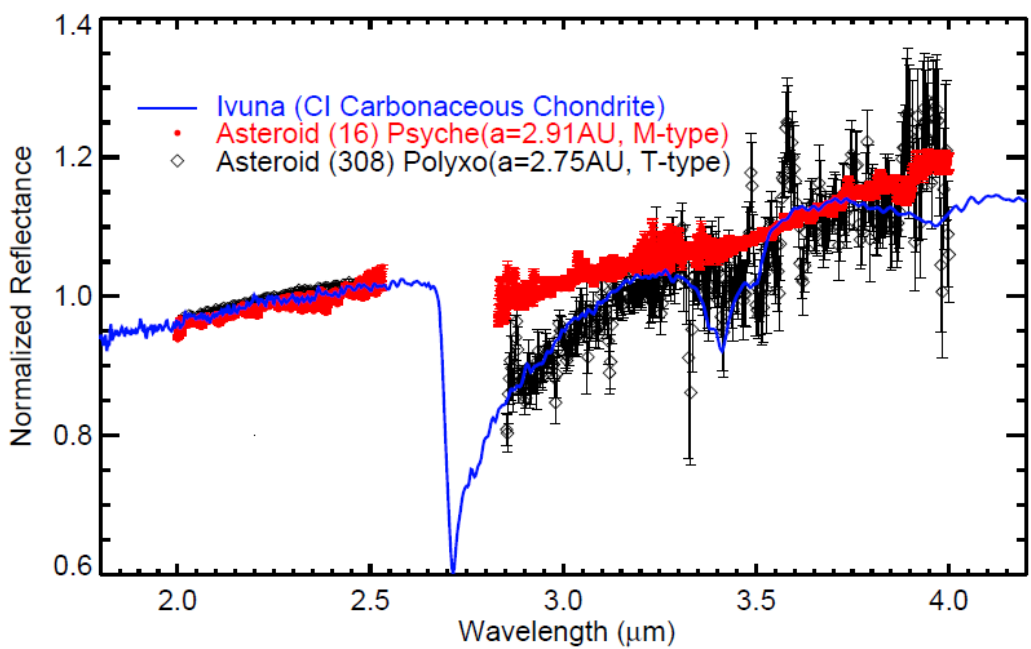

**Figure 5.** Psyche's spectrum is also found to be similar to the spectrum of the CI carbonaceous chondrite meteorite Ivuna (Takir et al. 2013) and T-type asteroid (308) Polyxo (a=2.75AU) (Takir and Emery 2012, Takir et al. 2015a). Both spectra of Ivuna and Polyxo have similar 3-µm band shape to the spectrum of Psyche with a deeper band depth (more abundance of hydrated minerals).

## 5. Summary

We observed asteroid (16) Psyche in the 3-µm band during two nights in December 2015, using (LXD: 1.9-4.2 µm) mode of the SpeX spectrograph/imager at the NASA IRTF. The LXD data were combined with the average NIR spectra and shape model of Psyche for deeper understanding of the asteroid. The observations of Psyche revealed that its spectra exhibit a 3-µm absorption features, possibly due to water or hydroxyl. The 3-µm band shape in Psyche's spectra was found to be consistent with the hydration features observed on some water-rich asteroids and carbonaceous chondrite meteorites, attributed to OH- and/or $H_2O$-bearing minerals. The detection of a 3-μm absorption band suggests that Psyche may not be metallic core, or it could be a metallic core that has been impacted by carbonaceous material over the past 4.5 Gyr. Our results also indicate rotational spectral variations, which we suggest reflect heterogeneity in the metal/silicate ratio on the surface of Psyche.

**Acknowledgements**

We wish to thank NASA IRTF staff for their assistance with asteroid observations. NASA IRTF is operated by the University of Hawai'i under contract NNH14CK55B with NASA. We are grateful for the thoughtful and excellent review by an anonymous reviewer. DT's contribution to this work was funded by Eugene M. Shoemaker Postdoctoral Fellowship (NASA Planetary Geology and Geophysics Program). Work by VR and JAS was funded NASA Planetary Geology and Geophysics grants NNX14AN05G and NNX14AN35G. JPE's contribution to this work was funded by NASA Solar System Observing grant NNX16AE91G.

# References

Asphaug, E., Agnor, C.B., Quentin, W. Hit-and-run planetary collisions. 2006. Nature 439, 155-160.Bus, S.I., Binzel, R.P., 2002. Phase II of the small Main-Belt asteroid spectroscopic survey: The observations. Icarus 158 (1), 146–177.

Bell, J.F.; et al. Asteroids- The big picture. In: Binzel, R.P., Gehrels, T., Matthews, M.S. (Eds.), University of Arizona Press, Asteroids II, pp. 921–945. 1989.

Bottke, W.F., Nesvorny, D., Grimm, R.E., Morbidelli, A., O'Brien, D.P. 2006. Iron meteorites as remnants of planetesimals formed in the terrestrial planet region. Nature 439 (7078):821-4.

Bowell, E., Hapke, B., Domingue, D., Lumme, K., Peltoniemi, J., Harris, A.W., 1989. Application of photometric models to asteroids. In: Binzel, R.P., Gehrels, T., Matthews, M.S. (Eds.), Asteroids II. The University of Arizona Press, Tucson, pp. 524–556. Bus, S.J., Binzel, R.P. Phase II of the small main-belt asteroid spectroscopic survey: A feature-based taxonomy. Icarus 158, 146–177.

Clark, R.N., 2009. Detection of Adsorbed Water and Hydroxyl on the Moon. Science 326, 562 doi:10.1126/science.1178105.

Cushing, M.C., 2004. Spextool: A Spectral Extraction Package for SpeX, a 0.8-5.5 Micron Cross-Dispersed Spectrograph. The Astronomical Society of the Pacific 116:362-376.

Davis, D.R., Farinella, P., Marzari, F., 1998. The missing Psyche family: Collisionally eroded or never formed? Icarus 137, 140–151.

DeMeo, F.E., Binzel, R.P., Slivan, S.M., Bus, S.J. 2009. An extension of the Bus asteroid taxonomy into the near-infrared. Icarus 202, 160–180.

De Santctis, M.C. et al. 2015. Ammoniated phyllosilicates with a likely outer Solar System origin on (1) Ceres. Nature 528, 241-244.

Farmer, C., The *Infrared Spectra of Minerals* (Mineralogical Society, monograph 4, London, 1974).

Fornasier S., Clark B. E., Dotto E., Migliorini A., Ockert-Bell M., and Barucci M. A. 2010. Spectroscopic survey of M-type asteroids. Icarus 210:655–673.

Gaffey M. J., E. A. Cloutis, M. S. Kelley and K. L. Reed (2002) "Mineralogy of asteroids. In Asteroids III" (W. F. Bottke, A. Cellino, P. Paolicchi and R. P. Binzel, Eds.), University of Arizona Press, pp. 183-204.

Harris, A.W., 1998. A thermal model for near-Earth asteroids. Icarus 131, 291–301.

Hardersen, P.S., Gaffey, M.J., and Abell, P.A. 2005. Near-IR spectral evidence for the presence of iron-poor orthopyroxenes on the surfaces of six M-type asteroids. Icarus 175:141-158.

Hardersen, P.S., Cloutis, E.A., Reddy, V., Mothe-Dinz, T., Emery, J.P. 2011. The M-/X-asteroid menagerie: Results of an NIR spectral survey of 45 main-belt asteroids. 2011. Meteoritics and Planetary Science 46:1910-1938.

Landsman, Z.A., Campins, H., Pinilla-Alonso, N., Hanus, J., Lorenzi, V. A new investigation of hydration in the M-type asteroids. Icarus 252. 186-198.

Lebofsky, L.A., Sykes, M.V., Tedesco, E.F., Veeder, G.J., Matson, D.L., Brown, R.H., Gradie, J.C., Feierberg, M.A., Rudy, R.J., 1986. A refined 'standard' model for asteroids based on observations of 1 Ceres and 2 Pallas. Icarus 68, 239–2.

Lupishko, D.F. On the bulk desnity of the M-type asteroid 16 Psyche. 2006. Solar System Research, Volume 40, Issue 3, pp 214-218.

Jones, T., Lebofsky, L., Lewis, J. & Marley, M. The composition and origin of the C, P, and D asteroids: Water as a tracer of thermal evolution in the outer belt.

Lebofsky, L. A. Asteroid 1 Ceres: Evidence for water of hydration. Mon. Not. R. Astron. Soc. 182, 17–21 (1978).

Magri, C., Ostro, S. J., Scheeres, D. J., Nolan, M.C., Giorgini, J.D., Benner, L.A.M., Margot, J.L. Radar observations and a physical model of Asteroid 1580 Betulia. Icarus 186: 152-177.

Matter, A., Delbo, M., Carry, B., Ligori, S. 2013. Evidence of a metal-rich surface for the Asteroid (16) Psyche from interferometric observations in the thermal infrared. Icarus 226, 419–427.

Milliken, R. E. and A. Rivkin (2009), Brucite and carbonate assemblages from altered olivine-r rich materials on Ceres, Nature Geoscience, 2, 258-261, doi:10.1038/NGEO478.

Muinonen, K., Belskaya, I.N., Cellino, A., Delbo, M., Levasseur-Regourd, A.C., Penttilä, A., and Tedesco, E.F. 2010. A three-parameter magnitude phase function for asteroids. Icarus 209:542-555.

Neeley, J.R., Clark, B.E., Ockert-Bell, M.E., Shepard, M.K., Conklin, J., Cloutis, E.A., Fornasier, S., Bus, S.j. 2014. The composition of M-type asteroids II: Synthesis of spectroscopic and radar observations. 2014. Icarus 238, 37-50.

Ockert-Bell, M.E., Clark, B.E., Shepard, M.K., Issacs, R.A., Cloutis, E.A., Fornasier, S., Bus, S.J., 2010. The composition of M-type asteroids: Synthesis of spectroscopic and radar observations. Icarus 210, 674–692.

Oszkiewicz, D.A., Muinonen, K., Bowell, E., Trilling, D., A. Penttilä, A., Pieniluoma, T., Wasserman, L.H., and Enga. M.T. 2011. Online multi-parameter phase-curve fitting and application to a large corpus of asteroid photometric data. Journal of Quantitative Spectroscopy and Radiative Transfer 112: 1919-1929.

Kuzmanoski, M. and Koračević, A. 2002. Motion of the asteroid (13206) 1997GC22 and the mass of (16) Psyche Astronomy & Astrophysics, 395, L17.

Reddy, V., Le Corre, L., O'Brien, D. P, et al. 2012. Delivery of dark material to Vesta via carbonaceous chondritic impacts. Icarus 221:544-559.

Rivkin, A.S. et al. 2000. The nature of M-class asteroids from 3-micron observations. Icarus 145 (2): 351. *Bibcode:2000Icar..145..351R. doi:10.1006/icar.2000.6354.*

Rayner, J.T. et al. 2003. SpeX: A medium-resolution 0.8–5.5-micron spectrograph and imager for the NASA Infrared Telescope Facility. The Astronomical Society of the Pacific 155:362-382.

Rayner, J.T., Onaka, P.M., Cushing, M.C., Vacca, W.D., 2004. Four years of good SpeX. The International Society for Optics and Photonics: 5492, 1498–1509.

Sanchez et al. 2016. Compositional characterization of asteroid (16) Psyche. American Astronomical Society Division for Planetary Science Meeting. Abstract# 325.20.

Shepard, M.K., and 19 colleagues, 2008. A radar survey of M- and X-class asteroids. Icarus 195, 184–205.

Shepard, M.K., Harris, A.W., Taylor, P.A., Clark, B.E., Ockert-Bell, M., Nolan, M.C., Howell, E.S., Magri, C., Giorgini, J.D., Benner, L.A.M., 2011. Radar observations of Asteroids 64 Angelina and 69 Hesperia. Icarus 215, 547–551.

Shepard, M.K. et al. 2016. In press. Radar observations and shape model of asteroid 16 Psyche. Icarus 000: 1-16.

Sunshine, J.M. Farnham, T.L., Feaga, L.M., Groussin, O., Merlin, F., Milliken, R.E., A'Hearn, M.F. 2009. Science **326**, 565 (2009). doi:10.1126/science.1179788.

Takir, D., Emery, J.P., 2012. Outer Main Belt asteroids: Identification and distribution of four 3-um spectral groups. Icarus 219, 641–654.

Takir, D. et al. 2013. Nature and degree of aqueous alteration in CM and CI carbonaceous chondrites. Meteorit. Planet. Sci. 48, 1618–1637.

Takir, D., et al. 2015a. Toward an understanding of phyllosilicate mineralogy in the outer main asteroid belt. Icarus. Icarus 257: 185-193.

Takir, D., et al. 2015b. Phase angle effects on 3-μm absorption band on Ceres: Implications for Dawn Mission. The Astrophysical Journal Letters, Vol 804, Number 1.

Taylor, J.R., 1982. An Introduction to Error Analysis. University Science Books, pp. 55–56.

Tholen, D.J., 1984. Asteroid taxonomy from cluster analysis of photometry. Ph.D. dissertation, University of Arizona.